\documentclass[twocolumn,preprintnumbers,showpacs]{revtex4-1}
\usepackage{amsmath}
\usepackage{amssymb}
\usepackage{graphicx}
\usepackage{color}
\usepackage{float}
\usepackage{sistyle}
\usepackage{subfigure}  
\usepackage{verbatim}   
\usepackage{graphicx}
\usepackage{dcolumn}
\usepackage{bm}
\usepackage[citecolor=blue,filecolor=blue,colorlinks=true,urlcolor=blue]{hyperref}
\usepackage{ulem}

\newcommand{\BAN}{\ensuremath{B_{1g}\,}}

\newcommand{\Ts}{\ensuremath{T^{\ast}\,}}
\newcommand{\Tc}{\ensuremath{T_{\rm c}\,}}

\newcommand{\cm}{\ensuremath{{\rm cm}^{-1}}}

\def\g{\gamma}
\def\w{\omega}

\graphicspath{{images/}{../images/}}

\begin{document}

\title{Vertical temperature-boundary of the pseudogap under the superconducting dome 
of the Bi$_{2}$Sr$_{2}$CaCu$_{2}$O$_{8+\delta}$ phase-diagram}

\author{B. Loret$^{1,2}$, S. Sakai$^3$, S. Benhabib$^1$, Y. Gallais$^1$, M. Cazayous$^1$, M. A. M{\'e}asson$^1$, R. D. Zhong$^4$, J. Schneeloch$^4$, G. D. Gu$^4$, A. Forget$^2$, D. Colson$^2$, I. Paul$^1$,  M. Civelli$^5$ and A. Sacuto$^1$ }

\affiliation{$^1$ Laboratoire Mat\'eriaux et Ph\'enom$\grave{e}$nes Quantiques (UMR 7162 CNRS), Universit\'e Paris Diderot-Paris 7, Bat. Condorcet, 75205 Paris Cedex 13, France\\
$^2$ Service de Physique de l'{\'E}tat Condens{\'e}, DSM/IRAMIS/SPEC (UMR 3680 CNRS), CEA Saclay 91191 Gif sur Yvette cedex France\\
$^3$ Center for Emergent Matter Science, RIKEN, 2-1 Hirosawa, Wako, Saitama 351-0198, Japan\\
$^4$ Matter Physics and Materials Science, Brookhaven National Laboratory (BNL), Upton, NY 11973, USA,\\
$^5$ Laboratoire de Physique des Solides, CNRS, Univ. Paris-Sud, Universit\'e Paris-Saclay, 91405 Orsay Cedex, France.}

\date{\today}


\begin{abstract}
Combining electronic Raman scattering experiments with cellular dynamical mean field theory, we present
evidence of the pseudogap in the superconducting state of various hole-doped cuprates. In Bi$_{2}$Sr$_{2}$CaCu$_{2}$O$_{8 + \delta}$ we track the superconducting pseudogap hallmark, a peak-dip feature, as a function of temperature $T$ and doping $p$, well beyond
the optimal one. We show that, at all temperatures under the superconducting dome, the pseudogap disappears at the doping $p_c$, between $0.222$ and $0.226$, where also the normal-state pseudogap collapses at a Lifshitz transition.
This demonstrates that the superconducting pseudogap boundary forms a vertical line in the $T-p$ phase diagram.

\end{abstract}


 \maketitle

Discovered thirty years ago \cite{Bednorz1986}, the copper oxide (cuprate) superconductors have not ceased to arise interest because their critical temperature \Tc is incredibly high at ambient pressure in comparison with conventional superconductors. Central to the high-\Tc cuprate problem is the challenge to understand the pseudogap (PG) state. In the normal phase, where the PG has been studied extensively, it manifests below a characteristic temperature \Ts $>$ \Tc as a loss of low energy spectral weight in spectroscopic responses \cite{Alloul1989, Warren1989,Homes1995,Fedorov1999, Opel2000, Fauque2006,Bernhard2008, Chatterjee2011, Vishik2012, Sacuto2013, Sakai2013,Benhabib15,Hashimoto2015,Mangin-Thro}, and indirectly in thermodynamical and transport properties \cite{Loram2001,Ando2004,Daou2009,Shekhter2013}. Its properties cannot be accounted for by the standard Fermi liquid theory of metals \cite{Abrikosov88,Nozieres1997}. 

An even greater challenge is to establish whether the PG exists in the superconducting phase, and if yes, what its doping dependence is. This is crucial to understand the relation between superconductivity and the pseudogap \cite{Anderson1987,Kotliar1988,Varma1999,Chubukov2008a,Moon2009}, which remains far from being well-understood \cite{Norman2005,Keimer2015}. However, there are only very few probes that can disentangle a pseudogap from a superconducting gap.
Note, even when the doping end-point of the normal state PG is known, it is unclear how that extrapolates in the superconducting phase, since it involves crossing a phase boundary. In the absence of an explicit method to identify the PG in the superconducting phase, this can be settled only through normal state extrapolations that require involved data analysis of heat capacity \cite{Loram2001} and angle-resolved photo-emission spectra (ARPES) \cite{Hashimoto2015}, or of magneto-resistivity and nuclear magnetic resonance measurements \cite{Alloul3,Daou2009,Badoux2016,Zheng2005,Kawasaki2010} under application of very high magnetic fields.

In this article, we present evidence that the PG develops in the SC state of different under-doped compounds, showing that it is a universal property of cuprates. In the case of Bi$_{2}$Sr$_{2}$CaCu$_{2}$O$_{8 +\delta}$ (Bi-2212), we are able to follow the PG evolution with doping under the superconducting dome. We show that the pseudogap end is a vertical line in the $T-p$ phase diagram within a narrow range of doping $0.222< p_c< 0.226$ \cite{Note2}, the doping level where a Lifshitz transition from a hole-like to an electron-like Fermi surface takes place in the underlying electronic structure \cite{Kaminski2006,Benhabib15}. Our experimental findings are analyzed within the cellular dynamical mean-field theory (CDMFT) applied to the two-dimensional Hubbard model.

\begin{figure}[tb!]
\begin{center}
\includegraphics[width=9cm,height=7cm]{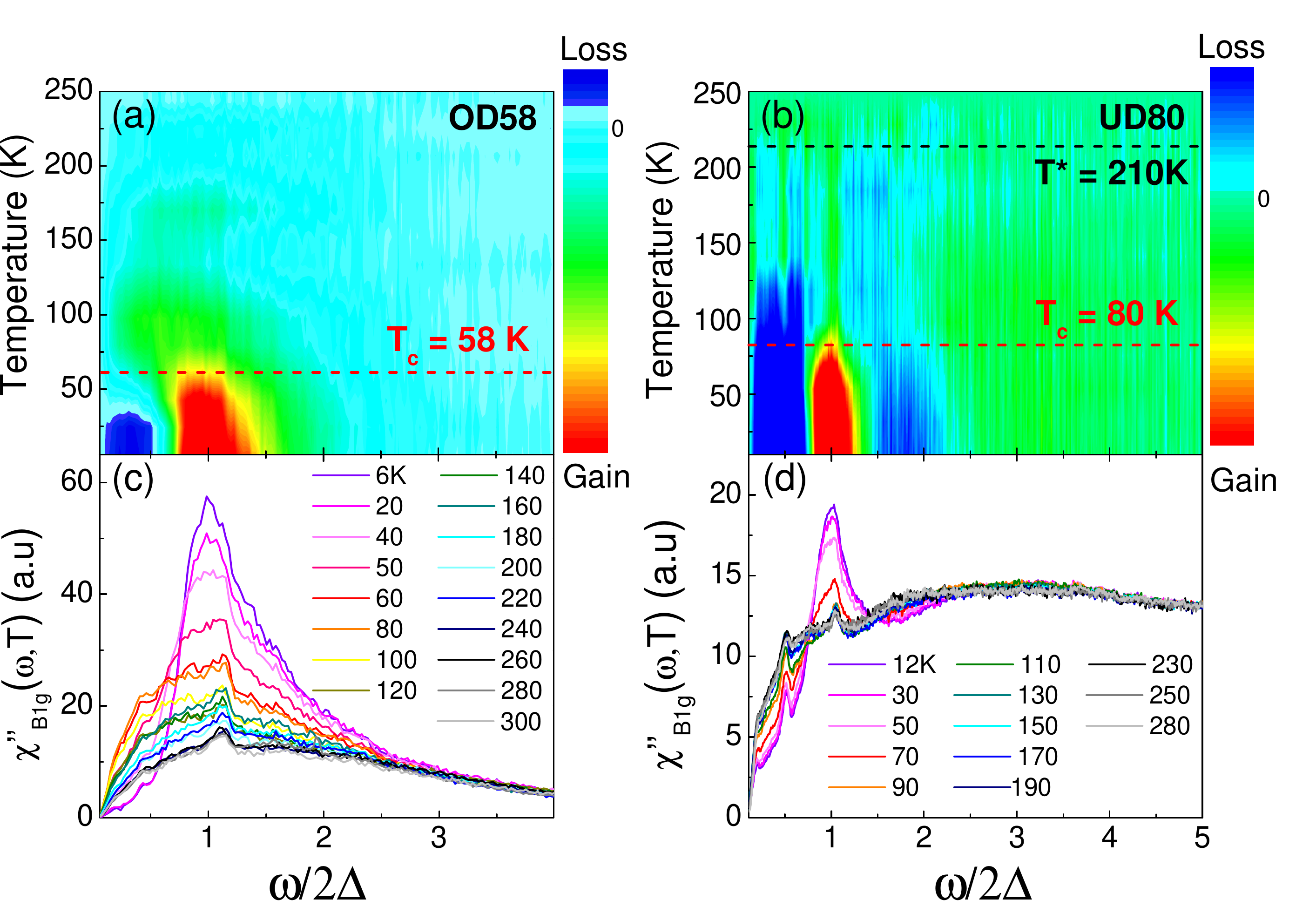}
\caption{(Color online). Cartographies of the \BAN Raman response difference $\chi(\omega, T)-\chi(\omega, T=250\, K$) versus temperature of (a) an over-doped (OD58, $p=0.226$) and (b) an under-doped (UD80, $p=0.123$) Bi-2212 single crystals. Cartographies were built from the \BAN Raman responses plotted in panels (c) and (d) subtracted from the ones at 250 K. The energy is expressed in $2\Delta$ units. $2\Delta$ corresponds to $247\,\cm$ and $558\,\cm$ for OD58 and UD80, respectively. The blue (dark gray) and the red (bright gray) colors correspond respectively to a loss and gain of Raman spectral weight.}
\label{fig:1}
\end{center}\vspace{-7mm}
\end{figure}
Before presenting our results, we shall explain how we identify the PG in the superconducting state. In this case, the PG manifests itself as a dip in the electronic background of the anti-nodal Raman response at frequencies higher than the pair breaking peak (PP). This PP-dip structure results from the interplay between the PG and the SC gap, and can be smoothly connected to the PG appearing in the electronic spectrum above \Tc \cite{Loret2016}.

In order to illustrate this point, we discuss the cartographies of the difference between the anti-nodal (\BAN) Raman response $\chi(\omega, T)-\chi(\omega, T=250\, K$) of an over-doped and under-doped Bi-2212 single crystals over a wide range of temperature and energy (Fig. \ref{fig:1}). Details of the Raman experiments, the crystal characterization and the doping setting are given in Supplementary Material (SM). 

We focus first on the overdoped compound (Fig.\ref{fig:1} (a)) for which there is no PG. At low temperature below \Tc, two distinct zones can be clearly seen: a blue (dark gray) zone at low energy representative of a loss of spectral weight and a red (gray) one around $2\Delta$ associated with the spectral weight transfer into the PP as expected when a SC gap is opening \cite{Klein1984,Devereaux2007}. The redistribution is partial because there is no sum rule in Raman scattering in contrast to that in optics \cite{Devereaux2007}. Above \Tc , this spectral weight redistribution disappears and there is no sign of the PG phase at this doping level.

On the other hand, the cartography of the under-doped compound (Fig.\ref{fig:1} (b)) is sharply different.
Below \Tc the PP is surrounded by two regions of spectral weight depressions (blue/dark gray), and spectral weight is transferred to the PP also from energies higher than $2\Delta$ \cite{Loret2016} leaving a spectral depression (blue/dark gray): This is the dip associated to PG in SC phase. Above \Tc, the dip (centered around $\omega/2\Delta \simeq$ 1.7) disappears while the loss of spectral weight below $\omega/2\Delta \simeq$ 1.5 persists and it merges with  the normal-state PG \cite{Opel2000,Venturini2002,Guyard2008,Sacuto2011,Sacuto2013}. The PP-dip feature is therefore the hallmark of the PG in the SC phase, which has been also confirmed theoretically by a CDMFT analysis on the Hubbard model \cite{Loret2016}. We note that in the normal phase the PG spectral depression completely vanishes nearby $210$ K (which corresponds to \Ts). A thorough analysis \cite {Note} not shown here, reveals this depletion is connected to a hump increasing in high energy electronic background (green/yellow zone around 3 of $2\Delta$), suggesting a transfer of spectral weight over a large frequency range as expected from CDMFT studies \cite{Sakai2013,Gull2013} and also detected by other studies \cite{Li2012,Bernhard2008}.

\begin{figure}[tb!]
\begin{center}
\includegraphics[width=8cm]{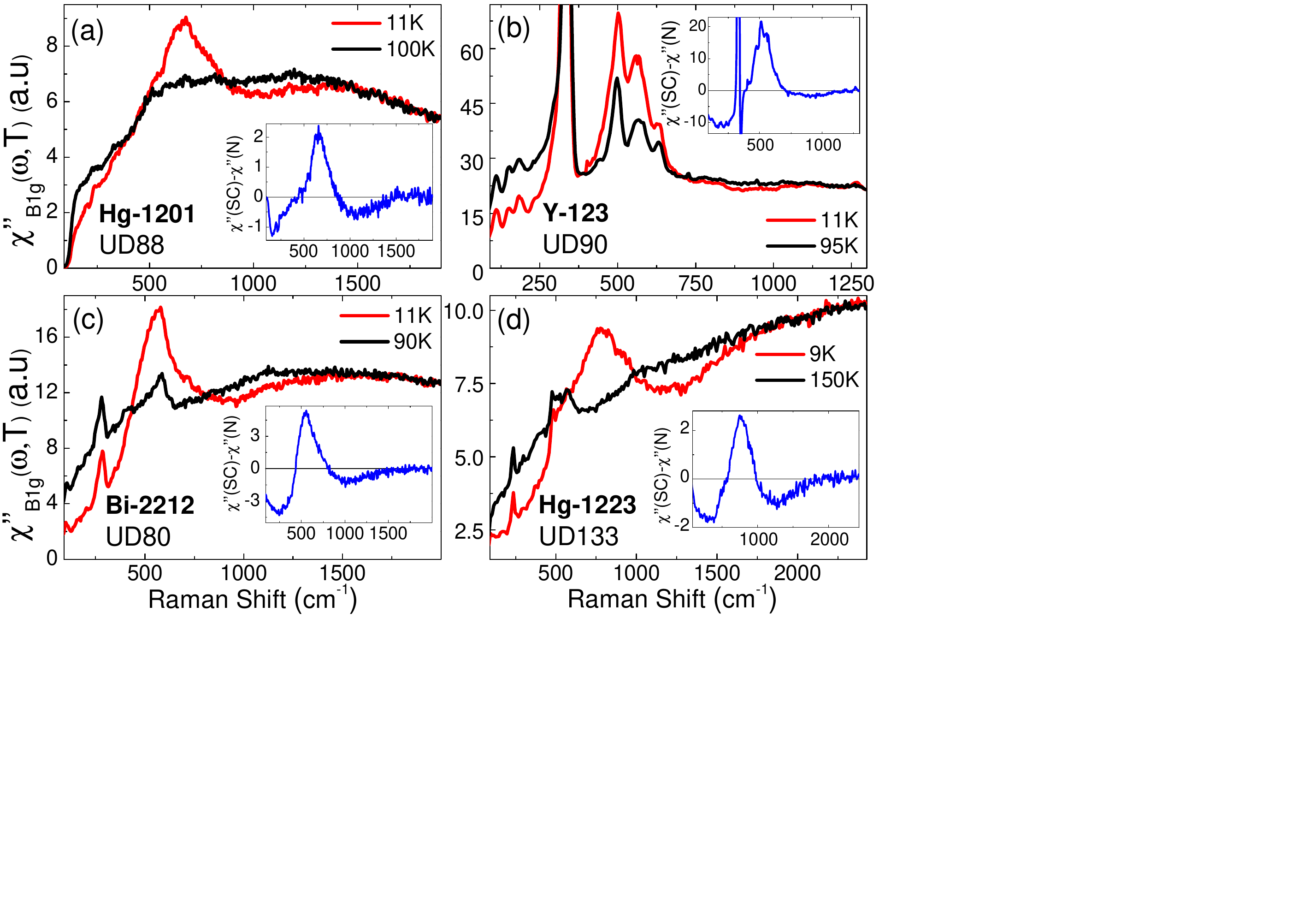}
\caption{(Color online). Superconducting and normal (just above \Tc) anti-nodal (\BAN) Raman responses $\chi^{\prime \prime}_{\BAN} (\omega, T)$ of four distinct underdoped cuprates (a) Hg-1201  (\Tc= $88$ K)  (b) Y-123  (\Tc=$90$ K), (c) Bi-2212  (\Tc=$80$ K), (d) Hg-1223  (\Tc=$133$ K). The Raman spectra in panels (b), (c) and (d) were obtained with the 532 nm laser line whereas the Hg-1201 one with the 647.1 nm line to reduce the Raman phonon activity. The PP-dip structure is also detected with the $532$ nm laser line in Hg-1201 showing the PP-dip structure is not a Raman resonant effect. Sharp peaks correspond to phonon modes. In the insets, a closer view of the PP-dip structure is plotted by subtracting the normal Raman spectrum from the superconducting one.}
\label{fig:2}
\end{center}\vspace{-7mm}
\end{figure}

Our first key result is that the PP-dip feature is present in several families of hole-doped cuprates, as shown in Fig.~\ref{fig:2}. Here the anti-nodal (\BAN) Raman responses $\chi^{\prime \prime}_{\BAN} (\omega, T)$ are displayed over a wide frequency range in the SC state and in the normal state (just above \Tc) for four distinct slightly under-doped cuprates: HgBa$_2$Cu$ $O$_{4+\delta}$ (Hg-1201), YBa$_{2}$Cu$_{3}$O$_{7-\delta}$ (Y-123), Bi$_{2}$Sr$_{2}$CaCu$_{2}$O$_{8+\delta}$ and HgBa$_2$Ca$_2$Cu$_3$O$_{8+\delta}$ (Hg-1223). The electronic background, which is superposed on various phononic peak contributions, strongly depends on the material. In order to disentangle the material-dependent features and put into evidence the universal character of PP-dip structure we have subtracted the normal-state Raman spectra (just above \Tc) from the superconducting ones as displayed in the insets of Fig.\ref{fig:2}(a)-(d).

The PP-dip feature is clearly observable at and above $2\Delta$ in all the compounds considered. The PP-dip disappears just above \Tc (see black curves) as reported in our previous work \cite{Loret2016}. The key observation here is that the presence of the PP-dip is independent of the number of copper-oxide layers: one layer (Hg-1201), two layers (Y-123 and Bi-2212) and three layers (Hg-1223), hence cannot be ascribed to a coupling between copper-oxygen planes \cite{Damascelli2003}.

\begin{figure}[tb!]
\begin{center}
\includegraphics[width=8.5cm,height=6.5cm]{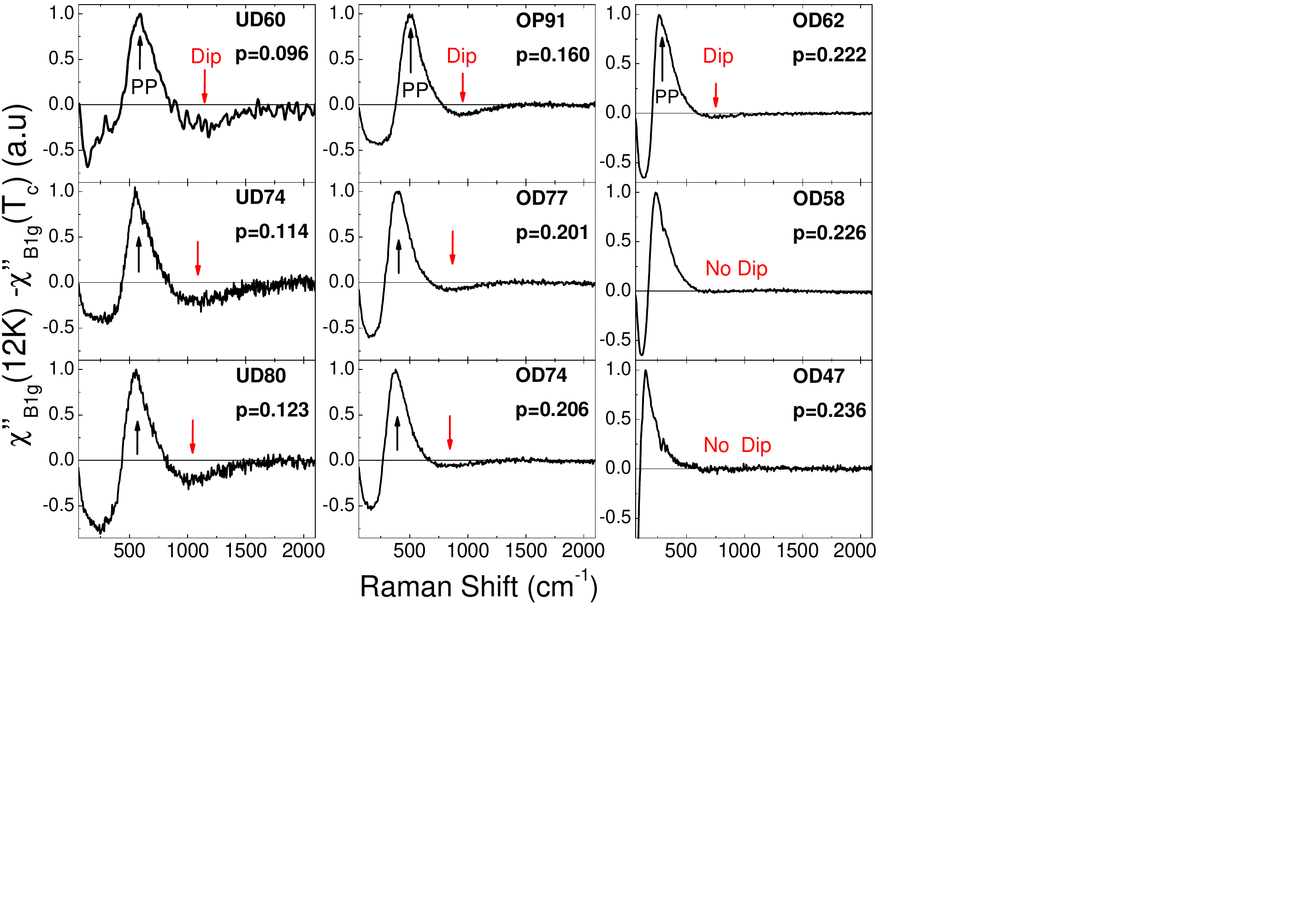}
\caption{(Color online).Difference of the antinodal (\BAN) Raman response in the SC ($12$K) and normal state (just above $T_c$) of Bi-2212 crystals for a set of doping levels from $p=0.096$ to $0.236$. The Raman intensity has been normalized to the maximum intensity of the PP. The dip disappears above $p=0.222$ while the PP is still observable. The arrows indicate the PP and the dip-bottom. Raw data are reported in the SM.}
\label{fig:3}
\end{center}\vspace{-7mm}
\end{figure}

We can use then the PP-dip as a hallmark to track the PG inside the SC dome with doping and find the doping level for which the pseudogap disappears. In Fig.~\ref{fig:3} we display the difference between the SC \BAN Raman responses of Bi-2212 at $12$ K and the one just above $T_c$  for several doping levels. The following are our main results concerning the $T-$doping evolution of the PP-dip feature.
(i) The PP-dip survives in the over-doped region. (ii) The dip, which is the PG feature in the superconducting phase, 
reduces with increasing doping. The continuation of the PG in the superconducting overdoped regime is in agreement 
with earlier works\cite{Zheng2005,Kawasaki2010,Vishik2012,Hashimoto2015}. (iii) Most interestingly, 
the PP-dip feature disappears in a narrow doping range between $p=0.222$ and $p=0.226$.

We now analyze how the PP-dip feature evolves as a function of doping comparing with the CDMFT results and examine if this gives a consistent picture. We calculate the Raman spectra of the two-dimensional Hubbard model with parameters appropriate for hole-doped cuprates: The (next-)nearest-neighbor transfer integral $t\sim0.3{\rm~eV}$ ($t'=-0.2t$) and the onsite Coulomb repulsion $U=8t$.
The CDMFT is implemented on a 2$\times$2 cluster and is solved with a finite-temperature extension of the exact diagonalization method \cite{Caffarel1994,Capone2007,Liebsch2012,Sakai2016a}.
Within the bubble approximation, the \BAN Raman response is calculated through
\begin{align}
\chi_{\BAN}''(\w)=2&\int \frac{d\mathbf{k}}{(2\pi)^2} \g_{\BAN} ^2(\mathbf{k})
 \int_{-\infty}^\infty d\w' [f(\w')-f(\w+\w')]\nonumber\\
&\times [{\rm Im}G(\mathbf{k},\w'){\rm Im}G(\mathbf{k},\w+\w')\nonumber\\
  &-{\rm Im}F(\mathbf{k},\w'){\rm Im}F(\mathbf{k},\w+\w')]
\label{eq:raman}
\end{align}
with $\g_{\BAN}$$=$$\frac{1}{2}[\cos(k_x)-\cos(k_y)]$ and $f(\w)$ being the Fermi distribution function.
Here, the normal ($G$) and anomalous ($F$) Green's functions calculated with the CDMFT are interpolated in the momentum space \cite{Kyung2006}. This approximation is quite robust in the anti-nodal region, which includes the cluster momenta $ \mathbf{K}= (0, \pm \pi)$, $(\pm \pi, 0) $, and will not affect our conclusions on the \BAN Raman response.
The 2$\times$2 CDMFT well portrays the richness of phases appearing in hole-doped cuprates, including the Mott insulator, the anti-ferromagnetism, the $d$-wave SC and the PG state\cite{Maier2005,Kotliar2006,Tremblay2006,Kancharla2008,Ferrero,Sordi2010,Gull2010,Gull2013,Gull2015}. In particular within CDMFT the PG originates from a singularity in the self-energy close to the Mott transition. This singularity evolves in the superconducting state, determining a prominent peak structure in the superconducting pairing function \cite{Sakai2016a,Sakai2016b} and, as outcome, the PP-dip structure in the Raman \BAN spectra \cite{Loret2016}. However, the optimal doping for which \Tc is maximal is $p^{th}_{opt} \approx 0.08 - 0.10$, which is smaller than the one ($p_{opt}\approx 0.16$) in experiments. A quantitative comparison with experiments is therefore not possible and we restrict ourselves to a qualitative one.

\begin{figure}[tb!]
\begin{center}
\includegraphics[width=8.5cm]{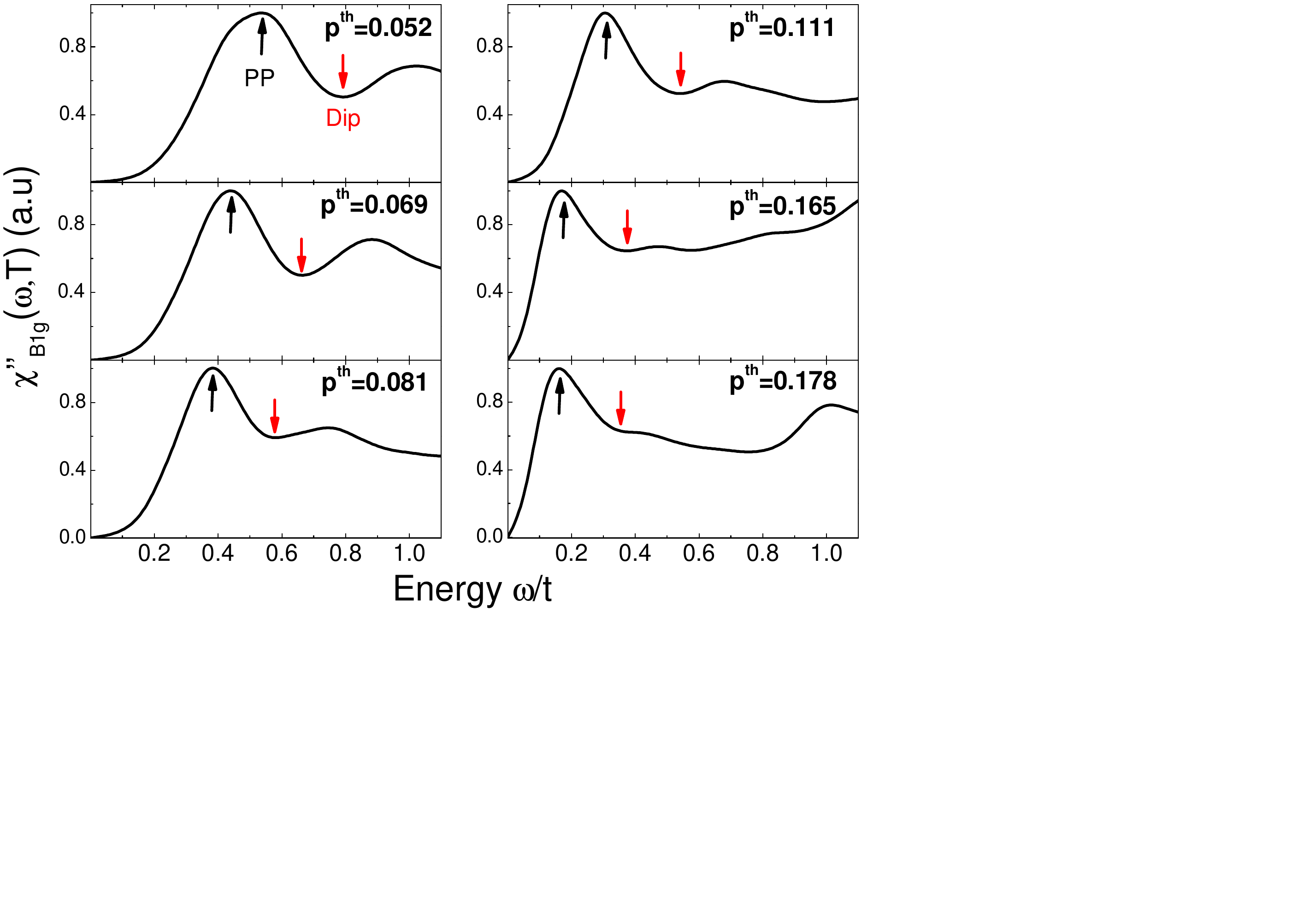}
\caption{(Color online). Theoretical \BAN Raman response obtained from Eq.(\ref{eq:raman}) combined with the CDMFT. The PP and the dip-bottom energies are denoted by arrows and go down with doping but their difference in energy remains almost constant with doping, as observed experimentally in Fig.3.} 
\label{fig:4}
\end{center}
\end{figure}

The CDMFT \BAN Raman responses in the SC state for increasing doping levels at a low temperature $T=0.005t$ are plotted on Fig.\ref{fig:4}.
Comparing with the experimental Raman spectra of Fig. \ref{fig:3}, we observe that the PP and dip trends are well reproduced by the CDMFT until $p^{th}=0.178$, which in the CDMFT phase diagram corresponds to a highly over-doped point. Both the theoretical and 
experimental dip-depth (pointed out by red arrows) reduce, while the distance in energy between the PP (black arrow) and dip-bottom (red arrows) stays roughly constant as a function of doping. At $p^{th}=0.178$, the theoretical dip, if still present, is very weak (see Fig. \ref{fig:4}). A Lifshitz transition is presumably located just above $p^{th}=0.178$ because the spectral function at the anti-nodal point is almost symmetric (see SM). Above $p^{th}=0.178$ the convergence of the CDMFT noticeably slows down and we could not obtain a well-converged solution. The closeness to the van Hove singularity might be one of the reasons why it becomes technically harder and harder to obtain converged CDMFT solutions. A more detailed comparison between experiment and theory will be published elsewhere.

\begin{figure}[t!!]
\begin{center}
\includegraphics[width=8.5cm,height=7cm]{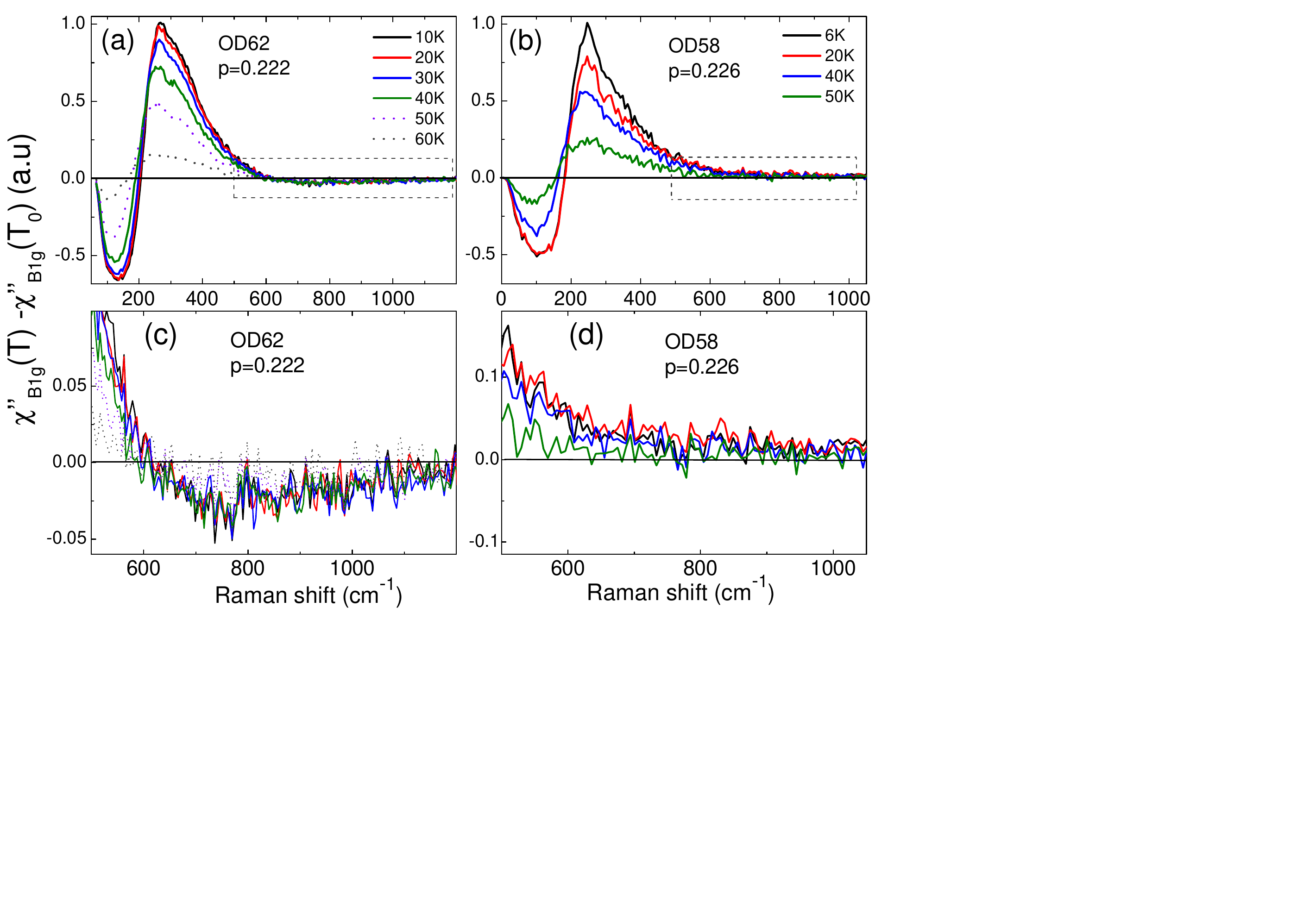}
\caption{(Color online). Temperature dependence of the difference between the SC \BAN Raman response and the one just above \Tc, denoted $T_0$  for (a) an overdoped OD62 Bi-2212 compound ($T_0= 70$ K), (b) an overdoped OD58 Bi-2212 compound ($T_0= 60$ K). (c) and (d) Closer views of the dip energy range above the pair breaking peak.}   
\label{fig:5}
\end{center} \vspace{-7mm}
\end{figure}

\begin{figure}[t!!]
\begin{center}
\includegraphics[width=8.5cm,height=7cm]{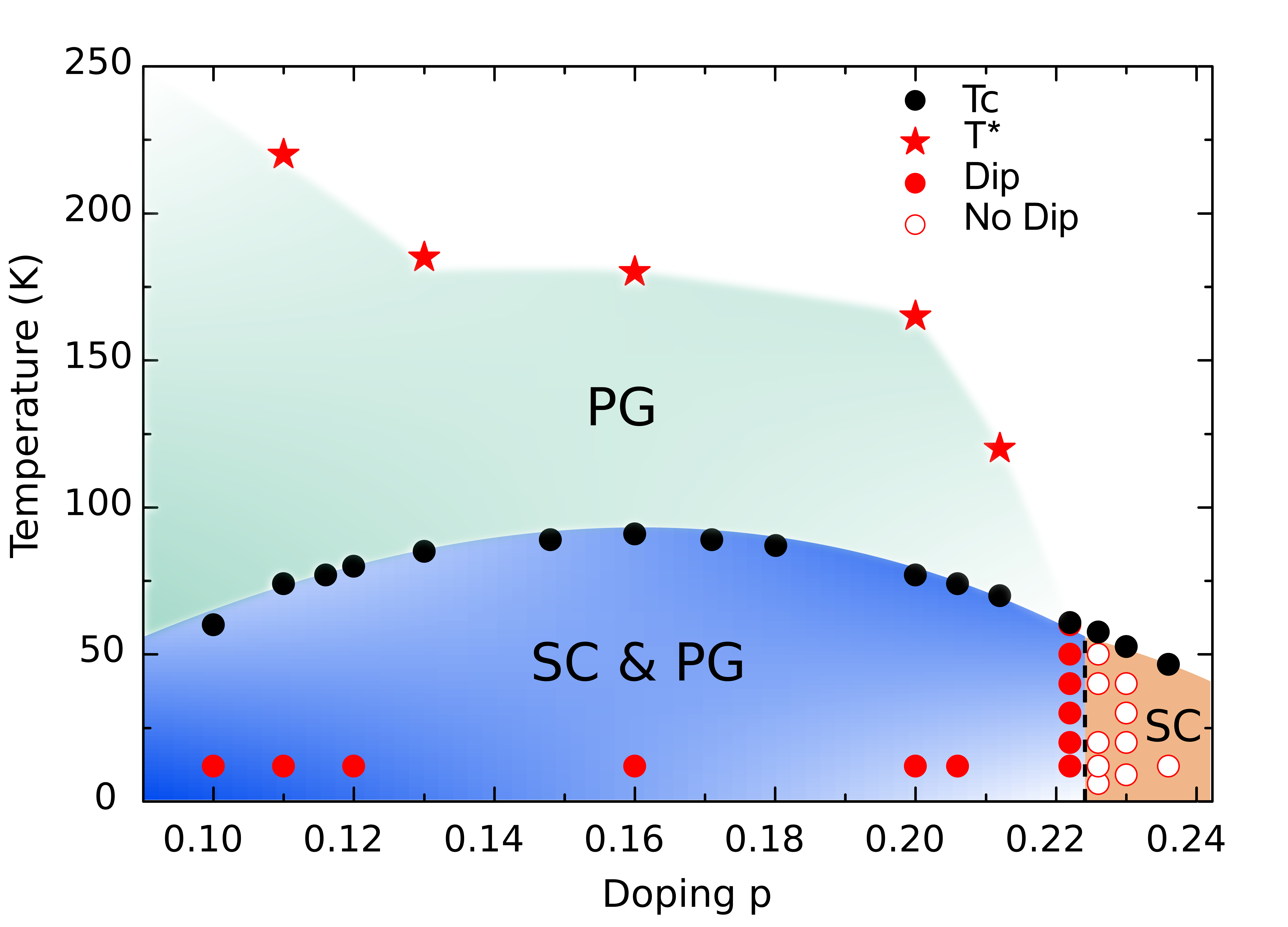}
\caption{(Color online). Temperature-doping phase diagram of Bi-2212, showing the PG in the normal and SC phases. The normal state PG which develops between \Ts and \Tc is obtained from the \BAN spectral loss observed in Ref.\cite{Benhabib15}. The \Ts values are reported from Ref.\cite{Benhabib15}. The dip is the PG-related feature in the SC state. The PG collapses abruptly (vertical line) between $p=$0.222 and $p=$0.226 in the SC state.}
\label{fig:6}
\end{center} 
\end{figure}

We finally show that at a doping $p_c$ which is between $p=$0.222 and 0.226, the PP-dip, and hence the PG, disappears from low temperatures ($T\approx10$ K) to \Tc. 
Fig. \ref{fig:5}(a) and (b) display respectively the $T-$dependent subtracted \BAN Raman response ($\chi"(T) - \chi"(T_0)$) of the OD62 ($p=0.222$) and OD58 ($p=0.226$) compounds, where $T_0$ is a temperature just above $T_c$, and is 70 K and 60 K, respectively. While the OD62 compound still displays a dip between 600 and $1200\, \cm$  for $T<$ \Tc, shown as a negative contribution in the close-up of Fig. \ref{fig:5}(c), the OD58 compound displays no dip over an equivalent temperature-range, as shown by the positive contribution in the closeup of Fig. \ref{fig:5}(d). This proves that the PG in Bi-2212 ends on a vertical line inside the SC dome of the $T-p$ phase diagram, which can be drawn between $p=0.222$ and $0.226$ (see Fig. \ref{fig:6}). 

Our result does not show any reentrant behaviour of the pseudogap inside the superconducting dome, in contrast
to that conjectured in \cite{Vishik2012,Hashimoto2015}, but rather a straight line, at least down to 12 K (well below that
of Ref.\cite{Hashimoto2015}). Note, our conclusions are based entirely on anti-nodal studies, unlike that of Ref.~\cite{Vishik2012}. 
Concerning the PG end-point in the normal state, our earlier results \cite{Benhabib15} are in good agreement with antinodal ARPES analysis\cite{Vishik2012,Hashimoto2015}. In this last case ARPES and Raman probe both antinodal quasiparticles. 

Our results strongly suggest that the superconducting PG, as in the normal state, is sensitive to the topology of the underlying Fermi surface since close to $p_c$ a Lifshitz transition takes place from a hole-like to an electron-like Fermi surface \cite{Benhabib15}. Furthermore, if the PG disappearance were a phase transition, it would be a first order one. This is expected for a Lifshitz transition of electrons coupled to a lattice \cite{Hackl2008,Kee2005}.

On the theory side, the relation between the pseudogap and the Lifshitz transition is not a well settled issue. The slowing down of the CDMFT solution approaching the van Hove doping level in concomitant with a strong decreasing of the dip depth is compatible with the experimental scenario, though future CDMFT improvements are needed to settle this issue.

In conclusion, we have shown that the peak-dip structure in the Raman \BAN spectra, which is the hallmark of the PG in the SC phase, is a universal feature of the hole-doped cuprates. Following the PP-dip evolution with doping and temperature in the case of Bi-2212, we show that the pseudogap persists on the over-doped side before disappearing abruptly and its end draws a vertical line in the $T-p$ phase diagram just in between $p=0.222$ and $p=0.226$. This corresponds to the same doping range where the normal-state pseudogap collapses, following up a Lifshitz transition of the Bi-2212 anti-bonding band, where the Fermi surface changes from holelike to electronlike.

We are grateful to A. Georges, A.J. Millis, S. Borisenko, and  Louis Taillefer for fruitful discussions. Correspondences and requests for materials should be addressed to A.S. (alain.sacuto@univ-paris-diderot.fr), M.C. (marcello.civelli@u-psud.fr) and  S.S  (shiro.sakai@riken.jp).  S.S. is supported by JSPS KAKENHI Grant No.~JP26800179 and JP16H06345; B.L. is supported by the DIM OxyMORE, Ile de France.

\newpage

\section{SUPPLEMENTAL MATERIAL}

\subsection*{Details of the Bi-2212 single crystal characterization}
The Bi-2212 single crystals were grown by using a floating zone method.~\cite{Wen2008,Mihaly1993}  
The critical temperature $T_{c}$ for each crystal has been determined from magnetization susceptibility measurements at a $10$ Gauss field parallel to the c-axis of the crystal. A complementary estimate of $T_{c}$  was achieved from electronic Raman scattering measurements by defining the temperature from which the \BAN superconducting pair breaking peak disappears. The level of doping $p$ was defined from $T_c$ using Presland and Tallon's equation\cite{Presland1991}: $1-T_{c}/T_{c}^{max} = 82.6 (p-0.16)^{2}$. In the over-doped regime, we have established a relationship between $T_c$ and the $2\Delta$ pair breaking peak. This give us another reliable way for directly estimate \Tc from $2\Delta$, and then evluate $p$ via the Presland and Tallon's equation, see details of section C in the SM of Ref. \cite{Benhabib15}.

\subsection*{Details of the Raman experiments}

Raman experiments have been carried out using a JY-T64000 spectrometer in single grating configuration using a 600 grooves/mm grating and a Thorlabs NF533-17 notch filter to block the stray light. The spectrometer is equipped with a nitrogen cooled back illuminated 2048x512 CCD
detector. Two laser excitation lines were used: 532 nm and 647.1 nm from respectively a diode pump solid state laser and a Ar+/Kr+ mixed laser gas. Measurements between 10 and 290 K have been performed using an ARS closed-cycle He cryostat. This configuration allows us to cover a wide spectral range (90 cm$^{-1}$ to 2200 cm$^{-1}$) in one shot. The \BAN-symmetry Raman response is obtained from crossed light polarizations along the Cu-O bond directions, giving us access to the anti-nodal region of the momentum space where the $d$-wave superconducting gap is maximal and the pseudogap sets in. All the spectra have been corrected for the Bose factor and the instrumental spectral response. They are thus proportional to the imaginary part of the Raman response function $\chi^{\prime \prime}(\omega,T)$ \cite{Sacuto2011,Sacuto2013}. In order to have reliable comparisons, all the Bi-2212 crystals have been measured in exactly the same experimental conditions. Special care has been devoted to cover a wide spectral range 
in one shot and maintain the laser spot at the same location on the crystal surface during the run in temperature. This makes reliable the intensities of the electronic Raman background from low to high energy without any adjusting the spectra from one temperature to another. We have checked that each spectrum is reproducible.

\subsection*{Raman responses of Bi-2212 in the superconducting and normal states}
\begin{figure}[ht!]
\begin{center}
\includegraphics[width=8cm,height=6cm]{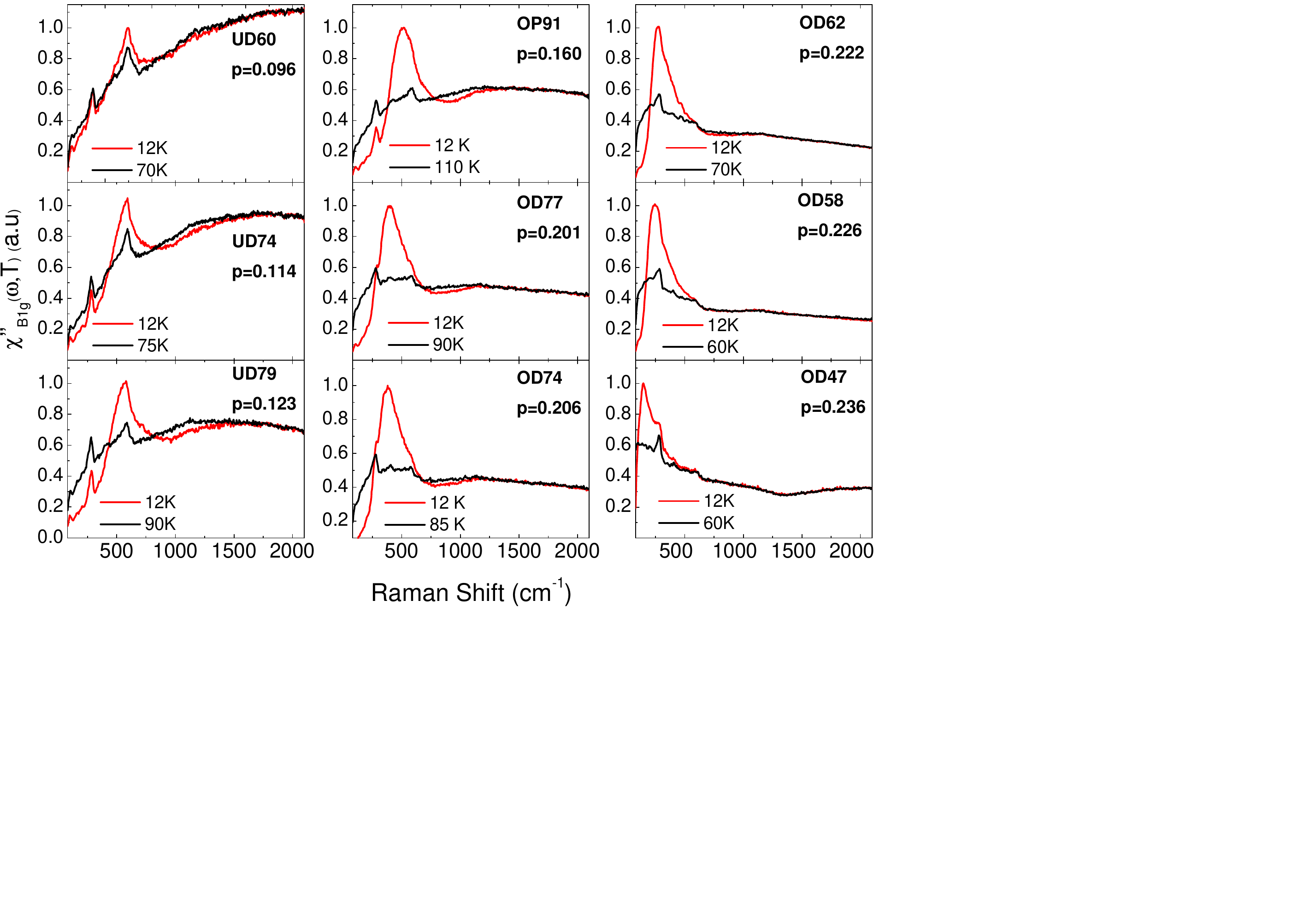}
\end{center}\vspace{-7mm}
\caption{Raman responses of Bi-2212 crystals normalized to the maximum intensity of the pair-breaking peak. All the spectra have been corrected for the Bose factor and the optical constants as determined by ellipsometry measurements \cite{Reznik2006}. The $532$ nm laser line was used for all the spectra.} 
\label{fig1S}
\end{figure}

In Figure \ref{fig1S} are reported the Raman responses of Bi-2212 single crystals for several doping levels in the superconducting ($T=12\,K$) and the normal state just above \Tc. The doping range extends from the under-doped to the over-doped regime. 

\subsection*{CDMFT spectral function}

\begin{figure}[ht!]
\begin{center}
\includegraphics[width=8cm]{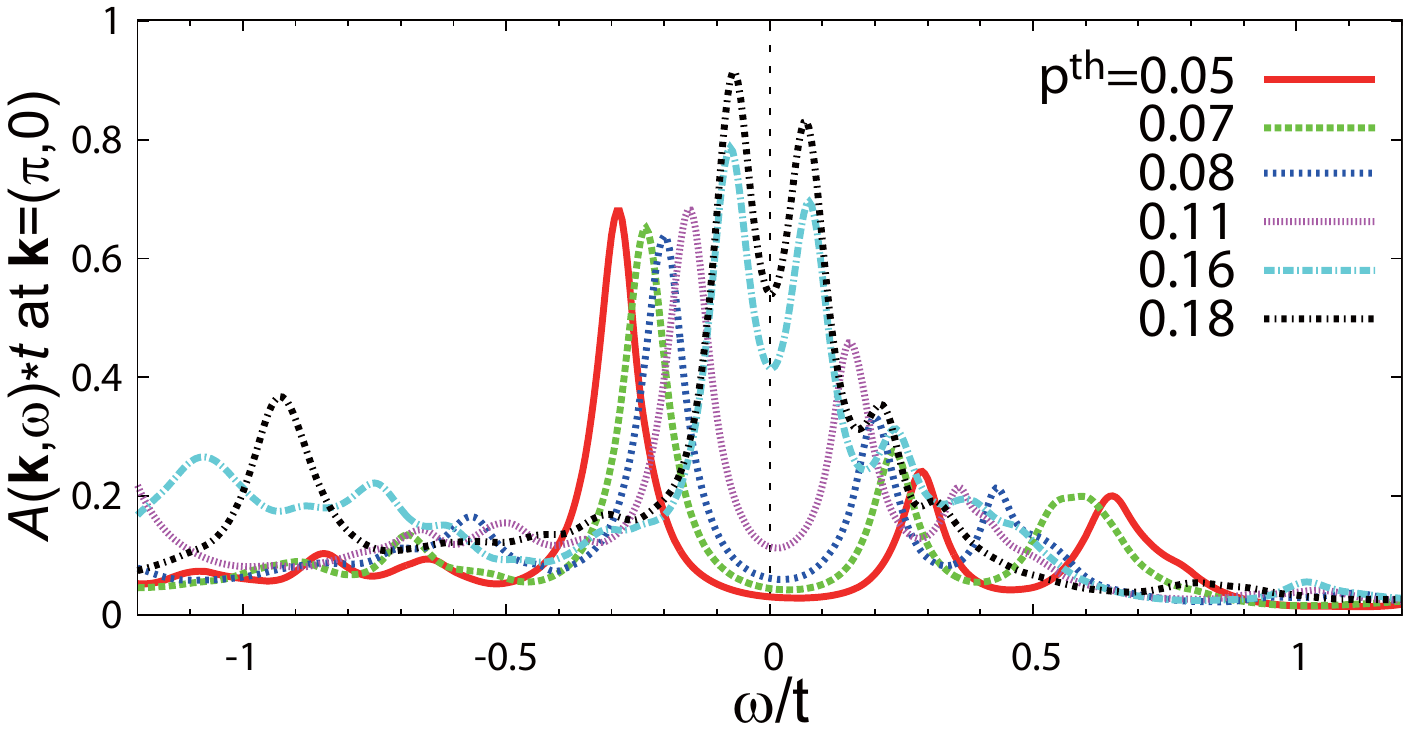}
\end{center}\vspace{-7mm}
\caption{Spectral function at $\textbf{k}=(\pi,0)$, calculated with the CDMFT for various doping levels.} 
\label{fig2S}
\end{figure}

In Figure \ref{fig2S} is plotted the spectral function $A(\textbf{k},\omega)$ at the antinodal point $\textbf{k}=(\pi,0)$, calculated with the CDMFT for various doping levels $p^{th}$.
The calculation was done at $T=0.005t$ in the superconducting state.
At small $p^{th}$, the spectra show a strong electron-hole asymmetry: The Bogoliubov peaks are much stronger for $\omega<0$ than $\omega>0$.
This asymmetry results from the underlying normal-state bare electronic structure, where the van Hove singularity is located below the Fermi energy.
As $p^{th}$ increases, the asymmetry decreases as expected.
Because the spectrum at $p^{th}=0.18$ is close to a symmetric one, we estimate that the Lifshitz transition point, where the underlying van Hove singularity crosses the Fermi energy, is located just above $p^{th}=0.18$.

\clearpage

\bibliographystyle{apsrev4-1}
\bibliography{referencesmain}

\begin{thebibliography}{64}
\bibitem [{Not({\natexlab{b}})}]{Note}%
  \BibitemOpen
  \href@noop {} {} ({\natexlab{b}}),\ \bibinfo {note} {a manuscript is in
  preparation.}\BibitemShut {Stop}%
\end{thebibliography}

\end{document}